\documentclass[3p]{elsarticle}
\usepackage{graphicx}
\usepackage{dcolumn}
\usepackage{xcolor}
\usepackage{bm}
\usepackage{array,multirow}
\usepackage{amssymb}
\usepackage{amsmath}
\usepackage{ulem}
\usepackage{lineno,hyperref}
\usepackage{makecell}
\modulolinenumbers[5]
\usepackage{comment}
\usepackage{soul}

\begin{document}

\begin{frontmatter}

\title{Scaling laws of elastic proton-proton scattering differential cross sections}

\author{Cristian Baldenegro \fnref{cbfootnote}}
\address{Dipartimento di Fisica, Sapienza Universit{\' a} di Roma, \\ Piazzale Aldo Moro, 2, 00185 Rome, Italy}
\fntext[cbfootnote]{c.baldenegro@cern.ch}
\author{Micha{\l} Prasza{\l}owicz\fnref{mpfootnote}}
\address{Institute of Theoretical Physics, Faculty of Physics, Astronomy and Applied Computer Science, \\
Jagiellonian University, S. {\L}ojasiewicza 11,
30-348 Krak{\'o}w, Poland.}
\fntext[mpfootnote]{michal.praszalowicz@uj.edu.pl}
\author{Christophe Royon\fnref{crfootnote}}
\address{Department of Physics and Astronomy, \\
The University of Kansas, Lawrence, KS 66045, USA}
\fntext[crfootnote]{Christophe.Royon@cern.ch}
\author{Anna M. Sta{\'s}to\fnref{asfootnote}}
\address{Department of Physics, Penn State University, \\University Park, PA 16802, USA.}
\fntext[asfootnote]{ams52@psu.edu}

\begin{abstract}
We show that elastic scattering $pp \to pp $ differential cross sections as function of the four-momentum transfer square $|t|$ have a universal property, such that the ratio of bump-to-dip positions is constant 
 from the energies of the ISR to the LHC, from tens of GeV and up to the TeV scale. We explore this property to compare the geometrical scaling 
 observed at the ISR with the recently proposed scaling law at the LHC. We argue that, at the LHC, within present
 experimental uncertainties, there is in fact a family of scaling laws. We discuss the constraints that the scaling 
 laws impose on the parametrization of the elastic $pp\to pp$ scattering amplitude.
 
\end{abstract}


\end{frontmatter}


\section{Introduction}

Recent precision measurements of the differential elastic $pp$
cross section by the TOTEM Collaboration at the LHC
\cite{TOTEM:2011vxg,TOTEM:2015oop,TOTEM:2017sdy,TOTEM:2018psk}, which led to
the discovery of the odderon~\cite{TOTEM:2020zzr}, has renewed interest in the phenomenological analyses of these highly non-perturbative observables. One expects that, at the LHC energies, the elastic scattering amplitude has a
much simpler structure than at the ISR
\cite{Nagy:1978iw,Amaldi:1979kd}. Indeed, the $t$-channel exchanges of the
colorless gluonic structures, namely the pomeron and the odderon, are
dominant at the LHC, whereas at lower energies exchanges of reggeons,  i.e., meson
exchanges, may also contribute. In both cases, the repulsive Coulomb
interaction is of importance at very low values of the four-momentum transfer squared $|t|$,
 see e.g. \cite{Kaspar:2011eva}.

The difference between the two energy ranges can be clearly seen by looking 
at the energy dependence of the
integrated cross-sections. At
the ISR, both elastic and inelastic total cross-sections have almost the same
energy dependence \cite{Amaldi:1979kd}, which means that their ratio is
approximately constant over the ISR energy range. This is no longer true at
the LHC~\cite{TOTEM:2017asr}.
We illustrate this by approximating the energy dependence of the elastic,
inelastic, and total cross-sections by power laws both at the ISR and the LHC energy
ranges separately. For the ISR fits we use data from Ref.~\cite{Amaldi:1979kd}
and for the LHC fits to TOTEM data from Ref.~\cite{Nemes:2019nvj}. Similar conclusions can be drawn by comparing with recent measurements by the ATLAS experiment~\cite{ATLAS:2014vxr,ATLAS:2022mgx}, 
modulo slight discrepancies between the TOTEM and ATLAS measurements. 
We use here only the TOTEM data for the LHC energies, since they cover a $|t|$ range sensitive to
relevant for the bump-and-dip ratio analysis of this paper, whereas the ATLAS Collaboration measurements have focused in the low $|t|$ region~\cite{ATLAS:2014vxr,ATLAS:2022mgx}. Uncertainties have been
estimated by changing the $\chi^2$ value by one unit.

The results are shown in Table \ref{tab:sigmas}. 
The most
interesting one is the energy dependence of the ratio $\sigma_{\mathrm{el}}%
/\sigma_{\mathrm{inel}}$, which, over the ISR energy range ($23.5-62.5$~GeV),
rises by approximately $0.5\%$, whereas for the LHC ($2.76-13$~TeV) the rise
is of approximately $13\%$. The fact that all ISR cross-sections scale
with energy approximately in the same way can be attributed, as will be
explicitly presented in Sec.~\ref{sec:GS@ISR}, to the (approximate) vanishment of the real part of the
scattering amplitude, as presented in Sec.~\ref{sec:GS@ISR}. At the LHC, all cross-sections scale differently, and
this is an important difference between the
two energy ranges.

\renewcommand{\arraystretch}{1.5} \begin{table}[h]
\centering
\begin{tabular}
[c]{|c|c|c|c|c|c|}\hline
& elastic & inelastic & total & $\frac{\mathrm{elastic}}{\mathrm{inelastic}}$
& $\rho$\\\hline
ISR & $W^{0.1142\pm0.0034}$ & $W^{0.1099\pm0.0012}$ & $W^{0.1098\pm0.0012}$ & $W^{0.0043\pm 0.0036}$ & $0.02-0.095$%
\\\hline
LHC & $W^{0.2279\pm0.0228}$ & $W^{0.1465\pm0.0133}$ & $W^{0.1729\pm 0.0163}$ & $W^{0.0814 \pm 0.0264}$ & $0.15-0.10$%
\\\hline
\end{tabular}
\caption{Energy dependence of the integrated cross-sections for the energies
$W=\sqrt{s}$ at the ISR \cite{Amaldi:1979kd}
and at the LHC \cite{Nemes:2019nvj} and the $\rho$ parameter \cite{Amaldi:1979kd,TOTEM:2017asr}.}
\label{tab:sigmas}%
\end{table}\renewcommand{\arraystretch}{1.0}

Differential elastic $pp$ cross-sections
also reveal  significant differences, even though the general structure is similar. 
One observes a rapid decrease for  small $|t|$, then a minimum at $t_{\rm d}$, called a {\it dip},
followed by a broad maximum at $t_{\rm b}$, dubbed as a {\it bump}.
It turns out, however, that the bump-to-dip cross-section ratio
\begin{equation}
\mathcal{R}_{\mathrm{bd}}(s)=\frac{  d\sigma_{\mathrm{el}} /d|t|_{\rm{b}}}
{ d\sigma_{\mathrm{el}} /d|t|_{\mathrm{d}}}\, ,%
\label{eq:Rbd}%
\end{equation}
which seems to saturate at LHC energies
 at a value of approximately 1.8,  is rather strongly energy
dependent at the ISR (see e.g. Fig.~2 in Ref.~\cite{TOTEM:2020zzr} and
Tab.~\ref{tab:dipbump}). This energy behavior of $\mathcal{R}_{\mathrm{bd}}$ 
is related to the change of relative importance of mesonic and gluonic contributions
to the elastic scattering amplitude, because meson exchanges vanish at high energies.

However, even for differential cross-sections there are some regularities that are common both to the ISR and the LHC.
The first one is  the
smallness of the real part of the forward elastic 
amplitude encoded in the
so called $\rho$ parameter, both at the ISR and the LHC energies, see Tab.~\ref{tab:sigmas}.  
In this Letter we explore yet another regularity of $ d\sigma_{\mathrm{el}} /d|t|$, which,
to the best of our knowledge, has not been used in phenomenological studies of the $pp$ elastic
scattering. Let us define the ratio of  bump-to-dip positions in $|t|$ at a given energy
\begin{equation}
\mathcal{T}_{\rm bd}(s)=|t_{\rm b}|/|t_{\rm d}|\, ,
\label{eq:Tbd}
\end{equation}
which, as we show below, is constant, within the experimental uncertainties, at all energies from the ISR to the LHC and equal approximately 
to 1.355. It is striking and unexpected that this ratio is constant over such large energy range.

The above mentioned property that bump/dip position ratio ${\cal T}_{\bm bd}$ (\ref{eq:Tbd}) is (almost) energy independent, suggests a universal
scaling variable both for the ISR and the LHC:
\begin{equation}
\tau= f(s) |t| \,\label{eq:taudef}%
\end{equation}
where $f(s)$ is a function of energy that can be found from fits to data.
Elastic differential cross-sections at different
energies, if plotted in terms of $\tau$, will have dips and bumps at exactly
the same values of $\tau_{\mathrm{d,b}}$. As already mentioned, at the LHC the values of cross-section ratios $\mathcal{R}_{\mathrm{bd}}$ (\ref{eq:Rbd}) are
(almost) energy independent, which again suggests a universal behavior
\begin{equation}
\frac{d\sigma_{\mathrm{el}}}{dt}(t_{\mathrm{d}})=g(s)\, \mathrm{const}%
_{\mathrm{d}},~~\frac{d\sigma_{\mathrm{el}}}{dt}(t_{\mathrm{b}})=g(s) \,
\mathrm{const}_{\mathrm{b}}\label{eq:xsecscaling}%
\end{equation}
where $g(s)$ has to be extracted from data. By analyzing the LHC data we extract both functions $f(s)$ and $ g(s)$. It is worth mentioning that recently an attempt to look for a scaling law at the LHC based on the saturation physics has been reported by some of us in Ref.~\cite{Baldenegro:2022xrj}. The energy behavior of the functions $f(s)$ and $g(s)$ is consistent with the findings of that work.

The observed scaling at the LHC has to be contrasted with the concept of the geometrical scaling proposed in the mid-seventies by Dias de Deus
\cite{DiasDeDeus:1973lde} in the context of the ISR data.  There, it was proposed that the scaling function has the form $f(s)=\sigma_{\mathrm{inel}}(s)$
\cite{Buras:1973km}. In addition, it was observed that rescaling $d \sigma
_{\mathrm{el}}/dt$ by $g(s)=\sigma_{\mathrm{inel}}^{2}(s)$ gives satisfactory
scaling except for the dip region. The crucial difference with respect to the LHC is that $\mathcal{R}_{\mathrm{bd}}$ at the ISR is energy
dependent. For other studies on scaling properties of elastic cross section see \cite{Dremin:2012qd, Csorgo:2019ewn}.

The structure of this Letter is the following. In the next section we discuss the structure of dip and bumps in the elastic cross section as function of $|t|$. By performing the fits to both structures, we extract their   positions in $|t|$ and find the universal behavior of their ratio. In Sec.~\ref{sec:GS@ISR}, we remind of the motivation of the geometrical scaling at the ISR by analyzing the cross-sections in impact parameter space. In Sec.~\ref{sec:GS@LHC}, the LHC published data are analyzed and the extraction of powers of energy  as functions $f(s)$ and $g(s)$ is performed. Sec.~\ref{sec:otherScalingLaws} is devoted to the comparison between the current analysis and scaling law proposed in Ref.\cite{Baldenegro:2022xrj}. In Sec.~\ref{sec:dipBump} we show the consequence of the scaling on the model of elastic amplitude which includes two components. Finally, in Sec.~\ref{sec:summary} we present summary and conclusions. 
\section{Dip and bump structure of the differential elastic cross-sections}

\begin{figure}[h]
\centering
\includegraphics[height=6.0cm]{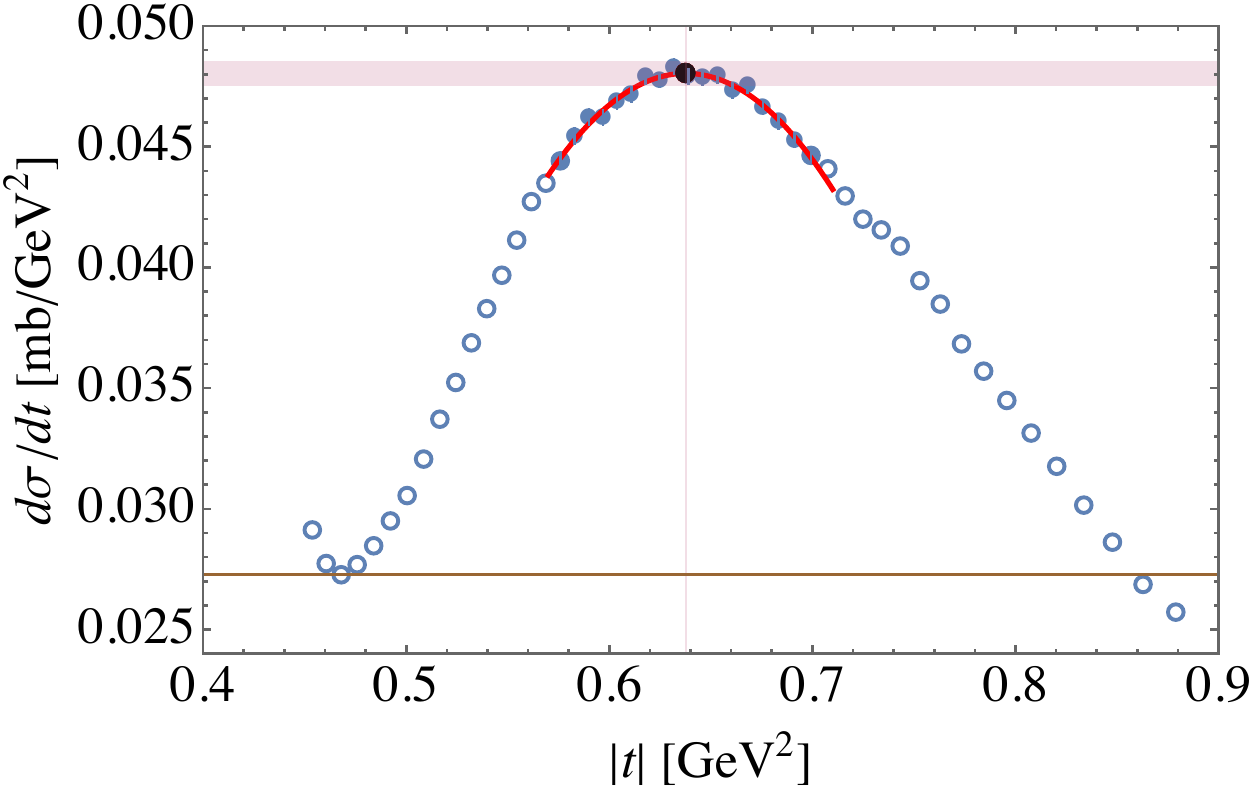}
\caption{Bump structure for the 13~TeV LHC data. From all the points above the solid line
chosen to cross the lowest point in the dip region, only a subset of them (in this case 18 points) are subject
to a parabolic fit. The top black point corresponds to the maximum and shaded areas  to the error bands of the position of such maximum as a result of the fit. The error band of the bump position in $|t|$ is very narrow, therefore not very well visible. Experimental errors are of the size of the point markers.}%
\label{fig:bump13}%
\end{figure}

In this Section, we describe the fitting strategy in the dip and bump regions both for the ISR and the LHC data. Although the
dip positions are relatively easy to find since the dip structure is always well pronounced, the
bump structure, on the contrary, is less well-defined due to the much flatter shape of the distribution in that region, such that  the quantitative determination of their positions is by far nontrivial. This is mainly due to the limited amount of LHC data (except 13~TeV), with large statistical and systematic uncertainties and/or a small number of data points in the dip or bump regions, depending on the data set. Ideally, one could first select bump or dip regions by drawing
a line through the lowest data point in the potential dip region, or through the highest point in the bump region,
and selecting all points above or below such line in a symmetric way, e.g., 5 points to the left and 5 points to the right with respect to a data point representing the maximum. This is illustrated on the example of the 13~TeV
LHC data set shown in Fig.~\ref{fig:bump13}, where in principle all data points above the solid brown line to the right
of a dip around 0.47~GeV$^2$  form a bump. We note that the 
13 TeV data set is the largest among the ones used here and is ideal for this type of analysis, and that the procedure described below is not fully feasible for other data sets in a uniform manner, as will be discussed later.

Once we have extracted a bump or dip region, we select a limited subset of points symmetrically around
a potential maximum/minimum, typically of the order of 10 to 15 data points. It is challenging to select the same number of points for all data sets, since the data sample sizes 
depend on the collision energy. A dedicated uncertainty is assigned to the arbitrariness of the selection of the number of points used in the fits, as explained later. In the case of 13 TeV data shown
in Fig.~\ref{fig:bump13}   the selected subset of, in this case, 18 points is
depicted as full circles. 
To this subset of points, we fit a parabola and the resulting bump point from such fit is shown in black in Fig.~\ref{fig:bump13}.
The statistical  uncertainty from the fit parameters is computed  from the condition that the $\chi^2$ of the fit changes by 1.
Of course, such a selection is arbitrary, but this is only observed for the 13 TeV data; in other cases it is difficult 
to find sometimes even seven points that qualify for a bump or dip. This is because,
as already discussed above,
the size of the experimental uncertainties of the LHC data is different at different energies, as well as the density of points and the range in $|t|$. Moreover, for the 8 TeV data, there is a gap in $|t|$ between 0.19 and 0.38 GeV$^2$, due to the use of two different beam optics configurations at the LHC.

Changing the number of data points subject to fitting symmetrically
upwards or downwards does not change the resulting fit function significantly (if we do not go too high or too low). To evaluate 
the systematic uncertainty associated to the arbitrariness of the fitting region, we remove one point from the left side of the maximum and add one point (that was previously not used in the fit) to the right side 
(and vice versa) and repeat the parabola fits (we dub this the ``asymmetric fits,'' since the maximum is the middle point). The difference between the symmetric and asymmetric fits defines the systematic uncertainties (left and right uncertainties may differ).  To estimate the total uncertainties, we add both uncertainties, 
statistical and systematic, in quadrature (shown as shaded areas in Fig.~\ref{fig:bump13}). We see from Fig.~\ref{fig:bump13}
that the uncertainty of the bump position  for $\sqrt{s}=13$~TeV is basically negligible.

\begin{figure}[h!]
\centering
\includegraphics[height=6.0cm]{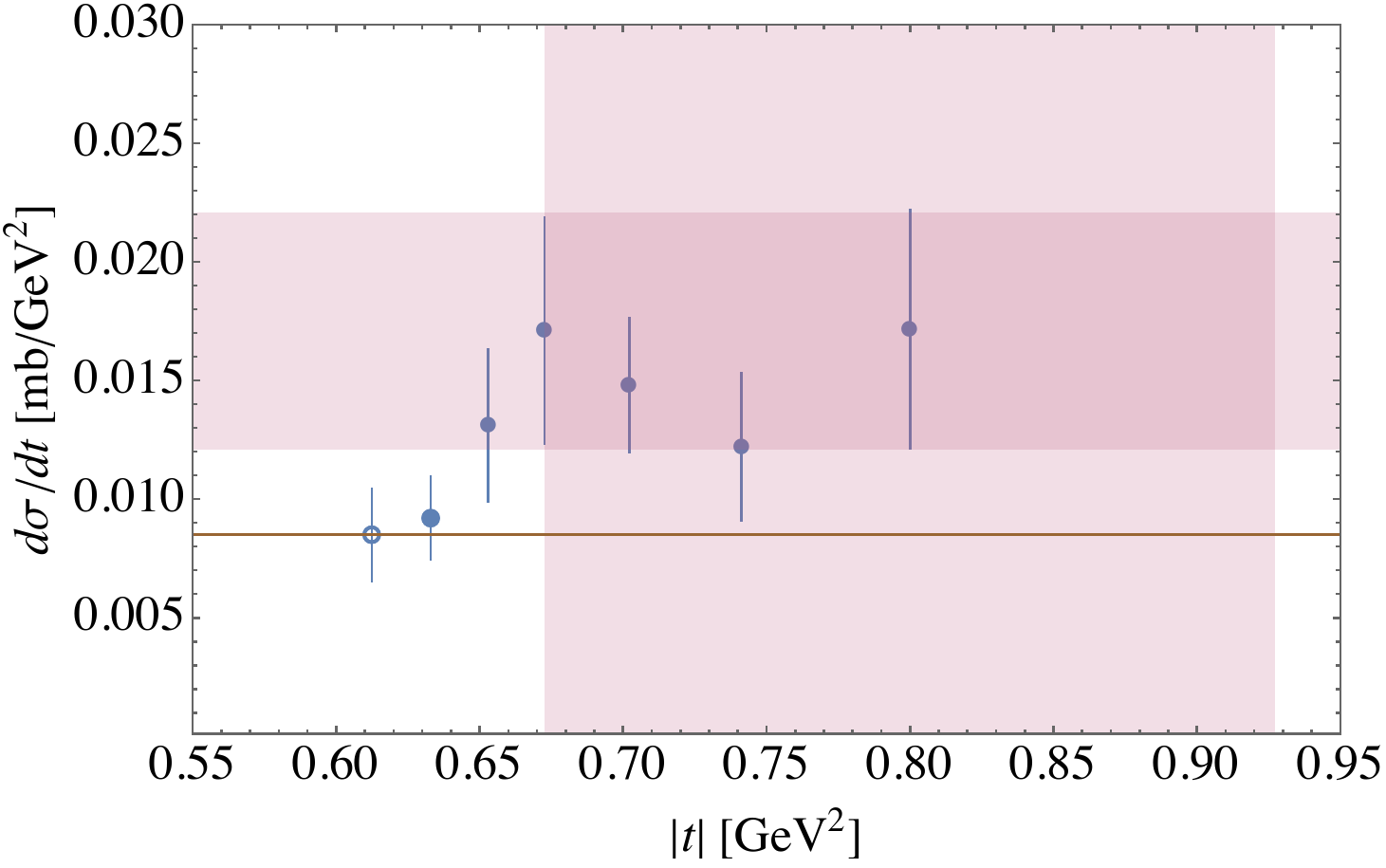}
\caption{Bump structure for the 2.76~TeV LHC data. In this case, most probably, the data points do not reach the underlying true bump.
Therefore, we choose the far right point as a potential candidate for the maximum, and the uncertainty is estimated as the distance
from this point to the highest point to the left. 
}%
\label{fig:bump276}%
\end{figure}

\begin{figure}[h!]
\centering
\includegraphics[height=5.58cm]{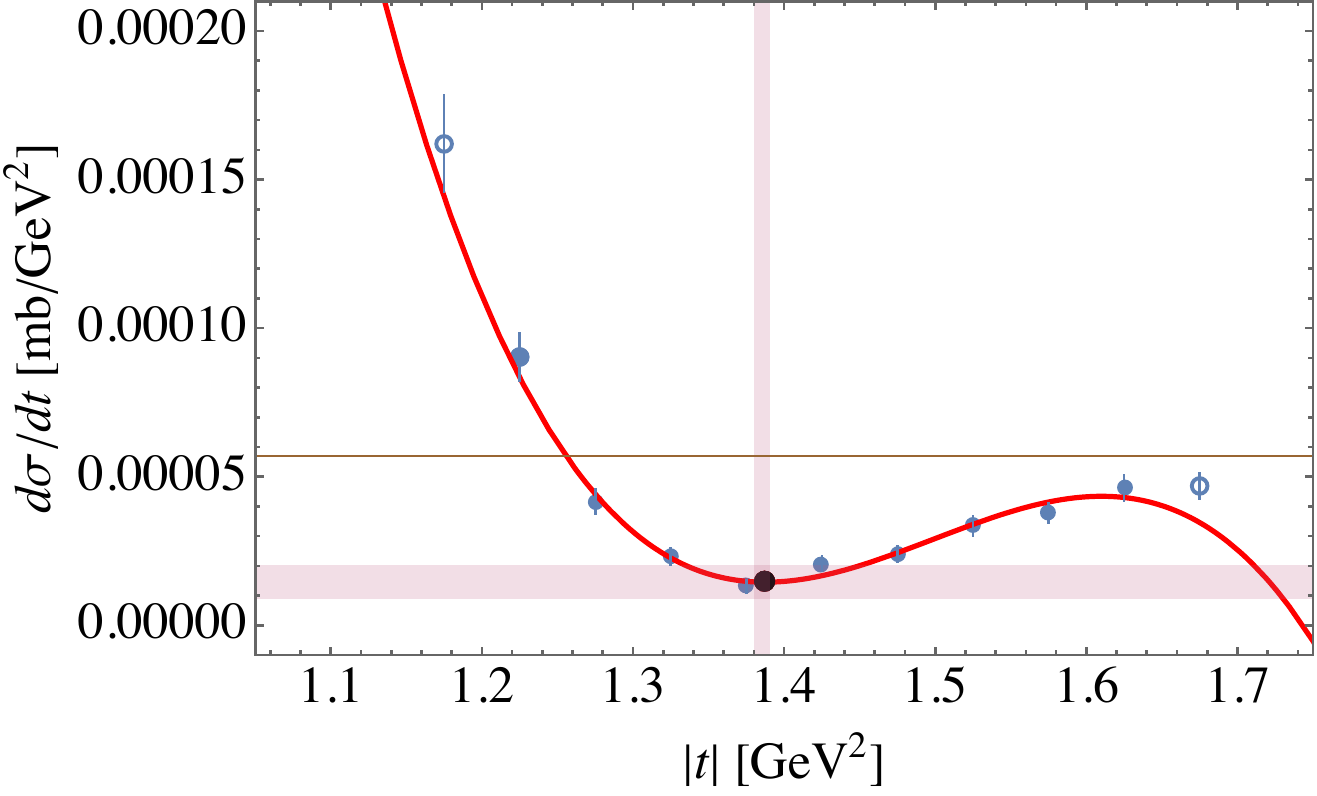}
\caption{Dip structure for the 40.64~GeV ISR data. Number of points below the solid line
that crosses the highest point in the bump region is small and the whole structure is clearly asymmetric.
Therefore, in this case we fit a third order polynomial rather than a parabola and include points above
the solid line left from the potential dip region. Open circles are for the points used to asses a systematic
uncertainty. Bottom black point corresponds to the minimum and shaded areas  to the error bands. The horizontal line corresponds to the value of the bump (as discussed at the beginning of Section~2) which is outside of the frame.}%
\label{fig:bump40}%
\end{figure}

The situation is quite  opposite in the case of the LHC data at 2.76 TeV. Here, in the bump region we have only 6 points with large uncertainties and, qualitatively, one can guess that the data might not have reached the true bump, see Fig.~\ref{fig:bump276}. Indeed, since it is well-established that the diffractive peak shrinks with
increasing energy, the bump of the 2.76~TeV data should be at a substantially larger  $|t|$ than approximately 0.7~GeV$^2$, which
is the bump position at 7~TeV. Therefore, we have assumed the bump to be at the far right with the systematic uncertainty given by the difference with the other
highest point at 0.673~GeV$^2$. The error of the maximum is assumed to be equal to the experimental error of the most right point.
This is shown in Fig.~\ref{fig:bump276}.

With the exception of 2.76~TeV, all bumps are reasonably well fitted by a parabola. Similarly, the dips at the LHC energies are fairly symmetric
and the parabola fits give  small enough $\chi^2$ values. On the contrary, dips in the case of the ISR data are strongly asymmetric and one had to fit a third-order polynomial to obtain good $\chi^2$, including one or two points from the diffractive peak above the line corresponding
to the bump, as shown in Fig.~\ref{fig:bump40} for the 40~GeV ISR data.

\renewcommand{\arraystretch}{1.3} 
\begin{table}[h!]
\centering
\begin{tabular}[c]{|c|c|r|lc|cl|cc|}\hline
\multirow{2}{*}{~}&\multirow{2}{*}{\#} & \multirow{2}{*}{$W$~} & \multicolumn{2}{c|}{dip} &\multicolumn{2}{c}{bump} & \multicolumn{2}{|c|}{ratios}\\
\cline{4-9}
& &  & $\left\vert t \right\vert _{\mathrm{d}} $ & error
&$\left\vert t \right\vert _{\mathrm{b}} $ & ~error
&$t_{\mathrm{b}}/t_{\mathrm{d}}$ & error
\\\hline
\multirow{4}{*}{\rotatebox{90}{LHC [TeV]}} 
& 9& 13.00 & 0.471 & $^{+0.002}_{-0.003}$ & ~0.6377 & $^{+0.0006}_{-0.0006}$ & 1.355 & $^{+0.008}_{-0.005}$\\
& 8& 8.00 & 0.525 & $^{+0.002}_{-0.004}$ & 0.700 & $^{+0.010}_{-0.008}$ & 1.335 &$^{+0.021}_{-0.016}$\\
& 7& 7.00 & 0.542 & $_{-0.013}^{+0.012}$ & 0.702 & $^{+0.034}_{-0.034}$ & 1.296 &$^{+0.069}_{-0.069}$\\
&6 & 2.76 & 0.616 & $^{+0.001}_{-0.002}$ & 0.800 & $^{+0.127}_{-0.127}$ & 1.298 &$^{+0.206}_{-0.206}$ \\
\hline
\multirow{5}{*}{\rotatebox{90}{ISR [GeV]}} 
& 5& 62.50 & 1.350 & $^{+0.011}_{-0.011}$ & 1.826 & $^{+0.016}_{-0.039}$ & 1.353 & $^{+0.016}_{-0.029}$\\
& 4& 52.81 & 1.369 & $^{+0.006}_{-0.006}$ & 1.851 & $^{+0.014}_{-0.018}$ & 1.352 & $^{+0.012}_{-0.014}$\\
& 3& 44.64 & 1.388 & $^{+0.003}_{-0.007}$ & 1.871 & $^{+0.031}_{-0.015}$ & 1.348 & $^{+0.023}_{-0.011}$\\
&2 & 30.54 & 1.434 & $_{-0.004}^{+0.001}$ & 1.957 & $_{-0.028}^{+0.013}$ & 1.365 &$_{-0.020}^{+0.010}$\\
& 1& 23.46 & 1.450 & $_{-0.004}^{+0.005}$ & 1.973 & $_{-0.018}^{+0.011}$ & 1.361 & $_{-0.013}^{+0.009}$\\
\hline
\end{tabular}
\caption{ Positions  of bumps  obtained
by fitting a parabola, and position of dips obtained by fitting a parabola (LHC) or a third order polynomial (ISR). Data set sequent numbers used in Fig.~\ref{fig:ratio} are listed in the second column. 
Dip positions at 13, 7 and 2.76~TeV agree with Ref.~\cite{Nemes:2019nvj}.}%
\label{tab:dipbump}%
\end{table}
\renewcommand{\arraystretch}{1.0}

\begin{figure}[h]
\centering
\includegraphics[height=6.5cm]{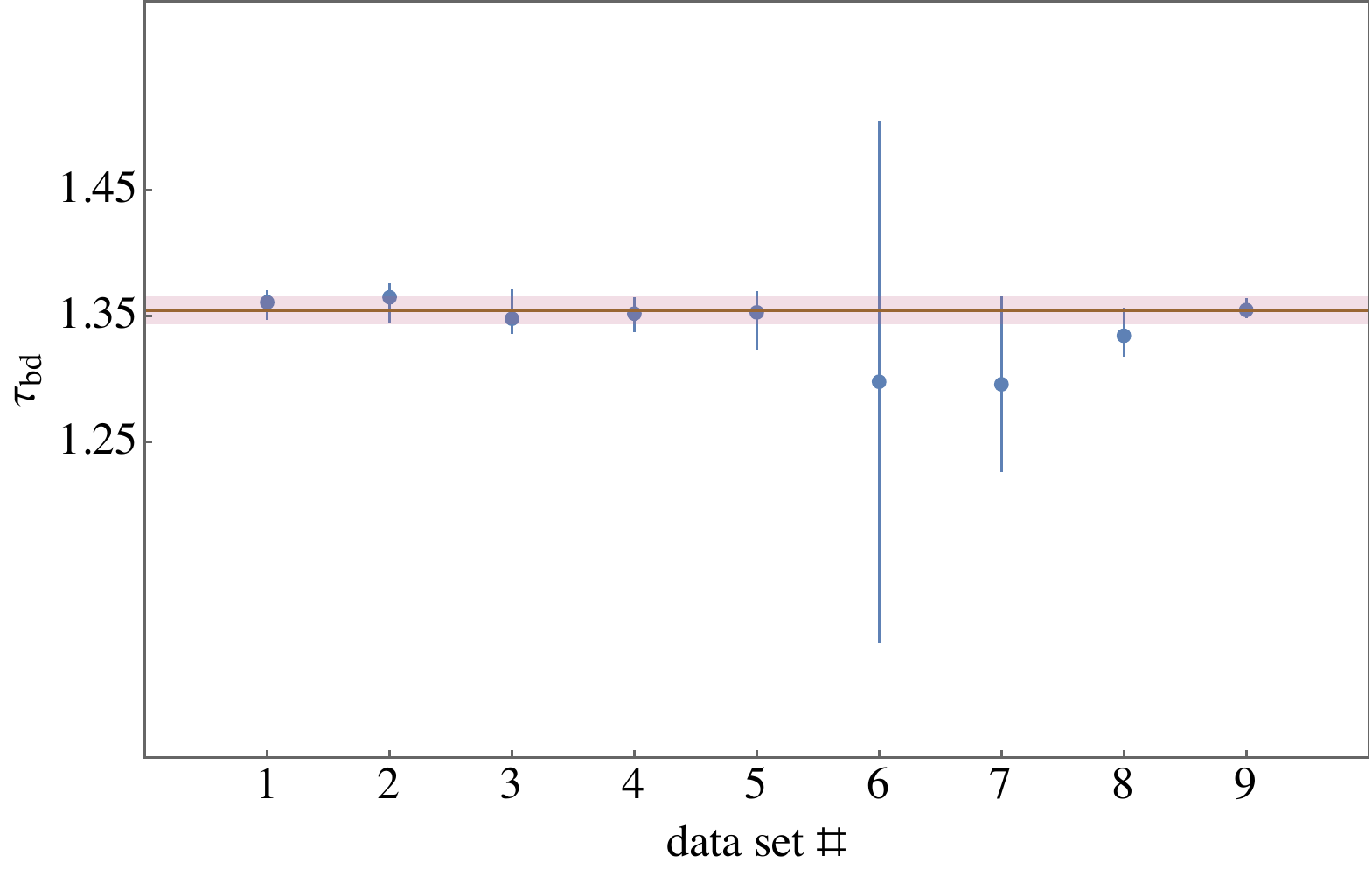}
\caption{Ratio $\mathcal{T}_{\rm bd}=|t_{\rm b}|/|t_{\rm d}|$ for the data sets from Tab.~\ref{tab:dipbump}.}%
\label{fig:ratio}%
\end{figure}

Numerical results obtained from the above procedure are collected in Tab.~\ref{tab:dipbump}. In the last column
we show the ratio of the bump/dip positions both for the ISR and LHC energies. We see that, unlike the ratio
of bump to dip cross-section values~\cite{TOTEM:2020zzr}, position ratio is energy independent and reads
\begin{equation}
\mathcal{T}_{\rm bd}=1.355\pm0.011 \, .
\label{eq:Tbd}
\end{equation}
We illustrate this in Fig.~\ref{fig:ratio}. 
Recall that at the LHC also the bump/dip cross-section ratios saturate, as discussed above.
In the following we shall
examine the consequences of these two properties at the LHC for geometrical scaling (GS)
of the elastic $pp$ cross-section. Before
that we shall recall the GS hypothesis developed in
the seventies for the ISR.

\section{Geometrical scaling at the ISR}
\label{sec:GS@ISR}

Unitarity constraints allow to write scattering cross sections in the impact
parameter space 
\begin{align}
\sigma_{\mathrm{el}}  & =%
{\displaystyle\int}
d^{2}\boldsymbol{b}\,\left\vert 1-e^{-\Omega(s,b)+i\chi(s,b)}\right\vert
^{2},\nonumber\\
\sigma_{\mathrm{tot}}  & =2%
{\displaystyle\int}
d^{2}\boldsymbol{b}\,\operatorname{Re}\left[  1-e^{-\Omega(s,b)+i\chi
(s,b)}\right]  ,\nonumber\\
\sigma_{\mathrm{inel}}  & =%
{\displaystyle\int}
d^{2}\boldsymbol{b}\,\left[  1-\left\vert e^{-\Omega(s,b)}\right\vert
^{2}\right]  \label{eq:sigmas}%
\end{align}
 where we expressed the impact parameter dependent amplitude in terms of the opacity  $\Omega(s,b)$ and phase $\chi(s,b)$, which are both real functions \cite{Barone:2002cv,Levin:1998pk}.
In Eqs.(\ref{eq:sigmas}) we have explicitly displayed possible phase
$\chi(s,b)$, which is responsible for the nonzero $\rho$ parameter
\cite{Levin:1998pk}. However, at the ISR energies the $\rho$ parameter is very
small~\cite{Amaldi:1979kd} (see Table \ref{tab:sigmas}), and one can safely
neglect $\chi(s,b)$ in the first approximation. Note that $\chi(s,b)$ does not enter the
expression (\ref{eq:sigmas}) for $\sigma_{\mathrm{inel}}$ \cite{Levin:1998pk}.

Geometrical scaling is a hypothesis such that%
\begin{equation}
\Omega(s,b)=\Omega\left(  b/R(s)\right) \label{eq:GSdef}%
\end{equation}
where $R(s)$ is the interaction radius \cite{DiasDeDeus:1973lde} increasing
with energy. Changing the integration variable from $\boldsymbol{b}%
\rightarrow\boldsymbol{B}=\boldsymbol{b}/R(s)$, one obtains that%
\begin{equation}
\sigma_{\mathrm{inel}}=R^{2}(s)%
{\displaystyle\int}
d^{2}\boldsymbol{B}\,\left[  1-\left\vert e^{-\Omega(B)}\right\vert
^{2}\right] \label{eq:siginel}%
\end{equation}
where the integral in (\ref{eq:siginel}) is an energy-independent constant. If
we neglect the phase $\chi(s,b)$, both elastic and total cross-sections should scale the same
way, which means that their ratios should be energy independent. As can be seen from Table~\ref{tab:sigmas}, this is indeed the case.

In Ref. \cite{Buras:1973km} from 1974, Buras and Dias de Deus confronted
the ISR data with the following scaling 
hypothesis
for the elastic $pp$ cross-section
\begin{equation}
\frac{1}{\sigma_{\mathrm{inel}}^{2}(s)}\frac{d\sigma_{\mathrm{el}}}%
{d|t|}(s,t)=\Phi(\tau)\label{eq:BDdDscaling}%
\end{equation}
where the scaling variable was defined as
\begin{equation}
\tau=\sigma_{\mathrm{inel}}(s)\,|t|\,=R^{2}(s)|t|\times\mathrm{const.}%
\label{eq:tau}%
\end{equation}
This relation follows from the expression for the differential elastic cross-section%
\begin{align}
\frac{d\sigma_{\mathrm{el}}}{d|t|}  & \sim\left\vert
{\displaystyle\int\limits_{0}^{\infty}}
db^{2}A_{\mathrm{el}}(b^{2},s)\,J_{0}\left(  b\sqrt{\left\vert t\right\vert
}\right)  \right\vert ^{2} 
\nonumber\\
& =\sigma_{\mathrm{inel}}^{2}(s)\left\vert
{\displaystyle\int\limits_{0}^{\infty}}
dB^{2}A_{\mathrm{el}}(B^{2})\,J_{0}\left(  B\sqrt{\tau}\right)  \right\vert
^{2}
\end{align}
where $A_{\mathrm{el}}(b^{2},s)$ is elastic scattering amplitude and $J_0$
denotes Bessel function originating from the Fourier transform.

\begin{figure}[h]
\centering
\includegraphics[height=7.5cm]{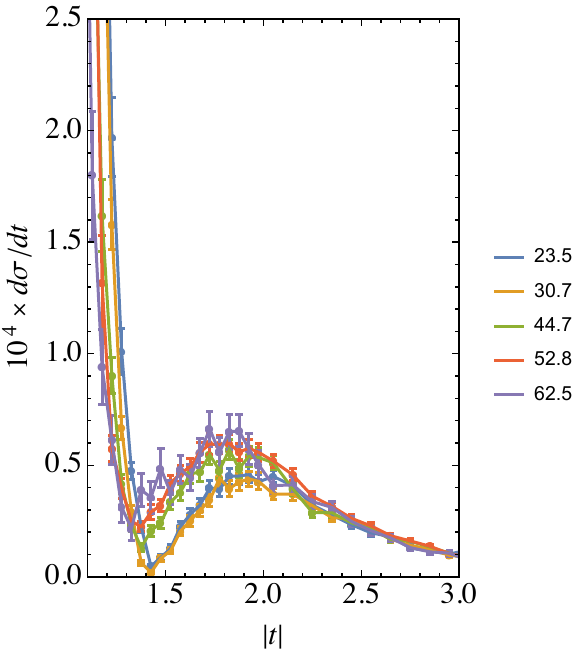}
~\includegraphics[height=7.5cm]{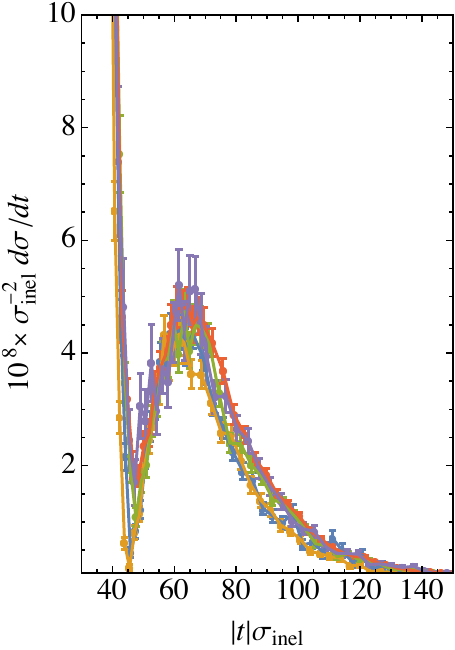}\caption{Elastic pp cross-section
$d\sigma_{\mathrm{el}}/dt$~[mb/GeV$^{2}$] multiplied by $10^4$
at the ISR energies in terms of
$|t|$~[GeV$^{2}$] -- left, and the same multiplied by $10^8$ 
and scaled according to
Eqs.~(\ref{eq:BDdDscaling},\ref{eq:tau}) -- right.
}
\label{fig:isr}%
\end{figure}

For the purpose of the comparison with the LHC, we have checked the
Dias de Deus -- Buras
scaling by replotting the ISR data in the form of (\ref{eq:BDdDscaling}).
The result is shown in Fig.~\ref{fig:isr}, where we display the dip-bump
region in linear scale.
The inelastic total cross-sections have been taken
from Ref.~\cite{Amaldi:1979kd}. We see that, indeed, scaling is clearly
visible, except in the dip region $t\sim t_{\mathrm{d}}$.  
The fact that geometric scaling is violated in the dip region was attempted to be explained by assuming that the imaginary part of the scattering amplitude vanishes at $t_{\mathrm{d}%
}$~\cite{DiasdeDeus:1977af}.
No matter how small  the real part of the amplitude, it 
would dominate in the
vicinity of $t_{\mathrm{d}}\pm\Delta t$. 
On the contrary for $|t-t_{\mathrm{d}}|>|\Delta t|$,
the imaginary part would dominate, and geometrical scaling  would be restored.

In order to visualize the quality of the scaling behavior,
we propose to use the {\it ratio method} that has been employed to assess the
quality of geometrical scaling in electron-proton deep inelastic scattering (DIS) in
Ref.~\cite{Praszalowicz:2012zh}. The method is based on an observation that if
some distributions (e.g., multiplicity distributions, differential
cross-sections, etc.) plotted in terms of a scaling variable $\tau$ coincide over some range of $\tau$, then
their ratio should be equal to unity in this range. To this end, we
choose the highest ISR energy $W_{\mathrm{ref}}=62.5$~GeV as a
reference energy  and construct ratios 
of the scaled cross-sections
\begin{equation}
R_{W}(\tau_{i})=\frac{\sigma^{-2}_{\mathrm{inel}}(W_{\mathrm{ref}}%
)\,d\sigma_{\mathrm{el}}/d|t|(W_{\mathrm{ref}},\tau_{i})} {\sigma
^{-2}_{\mathrm{inel}}(W)\,d\sigma_{\mathrm{el}}/d|t|(W,\tau_{i})} \,
.\label{eq:RW}%
\end{equation}

\begin{figure}[h]
\centering
\includegraphics[height=5.5cm]{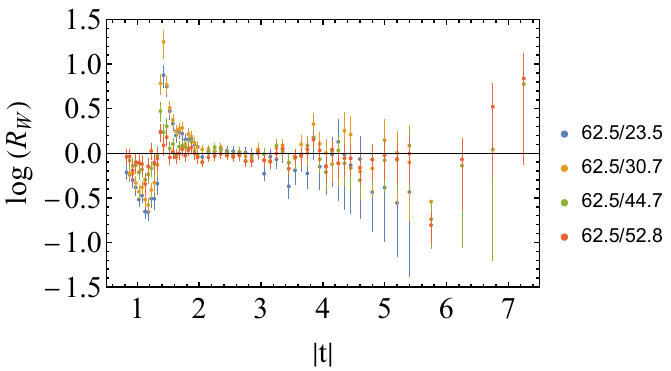} \\
\hspace{-2.cm}
\includegraphics[height=5.5cm]{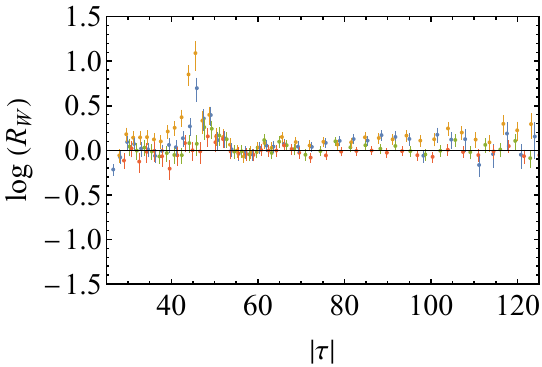}
\caption{Logarithm of ratios of differential
elastic $pp$ cross-sections at $W_{\mathrm{ref}}=62.5$~GeV to data at lower
energies ($W=52.8$~GeV -- orange, 44.7~GeV -- green, 30.7~GeV -- yellow and
23.5~GeV -- blue) before scaling (upper panel) and scaled (lower panel). }
\label{fig:isratios}%
\end{figure}

The disadvantage of this method is that we need to interpolate the
$W_{\mathrm{ref}}$ data to the points $\tau_{i}$, where the experimental data
for energy $W$ have been taken. To this end, we have performed  linear
interpolation between two consecutive points on both  the data and the uncertainty. We do not include
interpolation uncertainty in our analysis. The results are shown in
Fig.~\ref{fig:isratios} where we plot ratios for unscaled cross-sections (left
panel) and after scaling (right panel). We see that, after scaling,  $R_W$  approaches unity in a wide range in $|t|$, except in the vicinity of the dip where scaling is violated. Interestingly, we see scaling at very small values of $|t|$ below the dip region.

\section{Scaling at the LHC}
\label{sec:GS@LHC}

\begin{figure}[h]
\centering
\includegraphics[width=7.5cm]{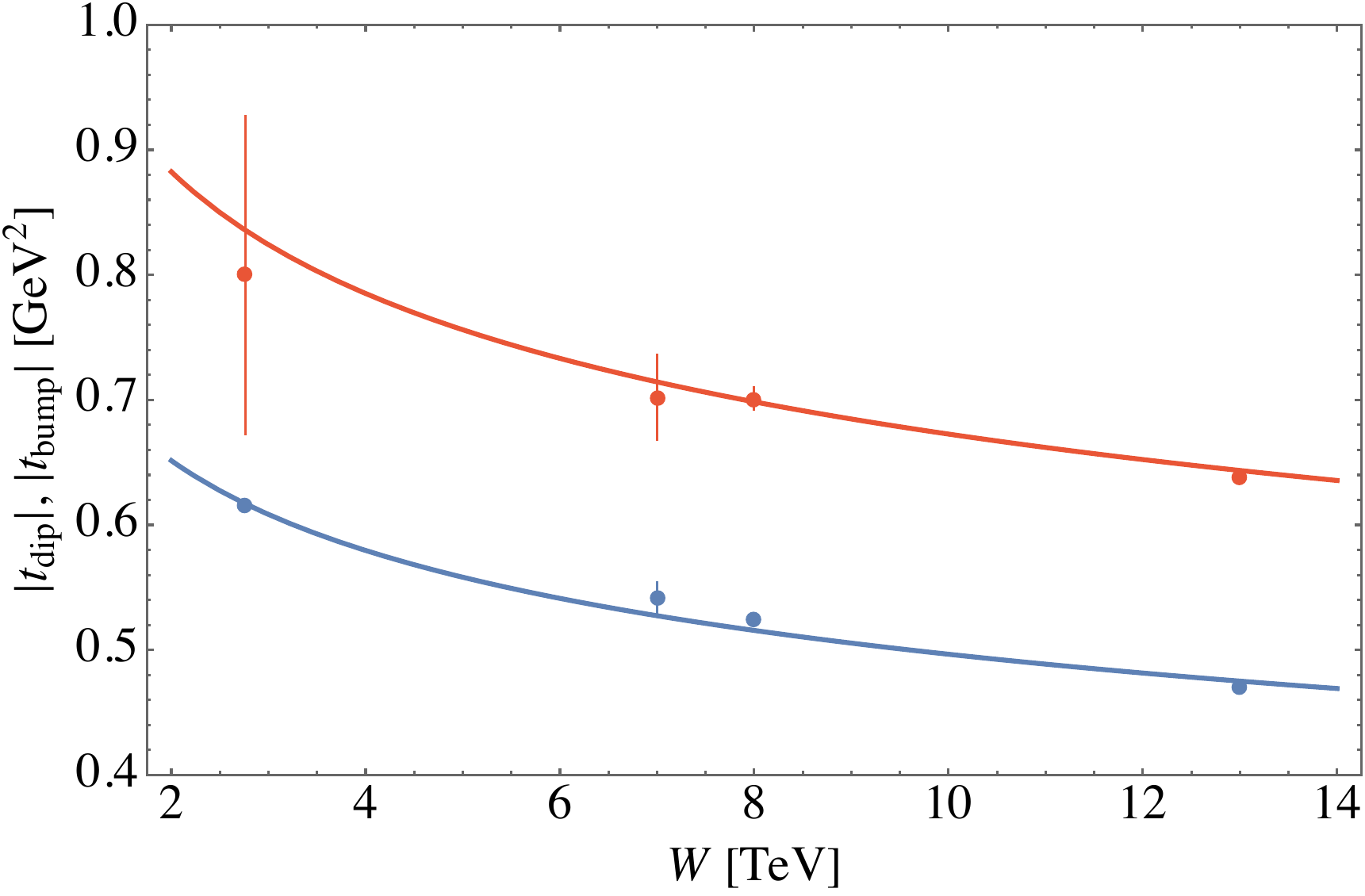} \vspace{-0.2cm}\caption{Fit to dip
and bump positions of Eq.~(\ref{eq:dipbumpfits}) with $\beta=0.1686$ at LHC energies.}%
\label{fig:dipfit}%
\end{figure}

In Ref.~\cite{TOTEM:2020zzr} the TOTEM data on elastic $pp$  differential cross-section were reanalyzed in search for the odderon. In
their analysis, they have identified a number of points of $d\sigma
_{\mathrm{el}}/d|t|$ called bumps, dips, and middle points, and plotted them in terms of
the scattering energy $W=\sqrt{s}$ (see Fig.~3 in Ref.~\cite{TOTEM:2020zzr}).
In what follows, we assume that both dip and bump positions can be fitted
with a power law function, i.e. $f(W)=B\, W^{-\beta}$ in Eq.~(\ref{eq:taudef}). However,
due to large uncertainties on the bump positions (see Fig.~\ref{fig:dipfit}), we have performed
only a fit to the dips and for the bumps we have used the result of Eq.~(\ref{eq:Tbd}):
\begin{equation}
t_{\mathrm{dip}}(W)=(0.732\pm 0.003)\times(W/(1~\mathrm{TeV}))^{-0.1686\pm 0.0027}
~~~{\rm and}~~
t_{\mathrm{bump}}(W)=1.355 \times t_{\mathrm{dip}}(W) \, .
\label{eq:dipbumpfits}%
\end{equation}
We have checked that $t_{\mathrm{bump}}(W)$ of Eq.~(\ref{eq:dipbumpfits})
gives  $\chi^2\approx 1$.

\begin{figure}[h]
\centering
\includegraphics[height=6.0cm]{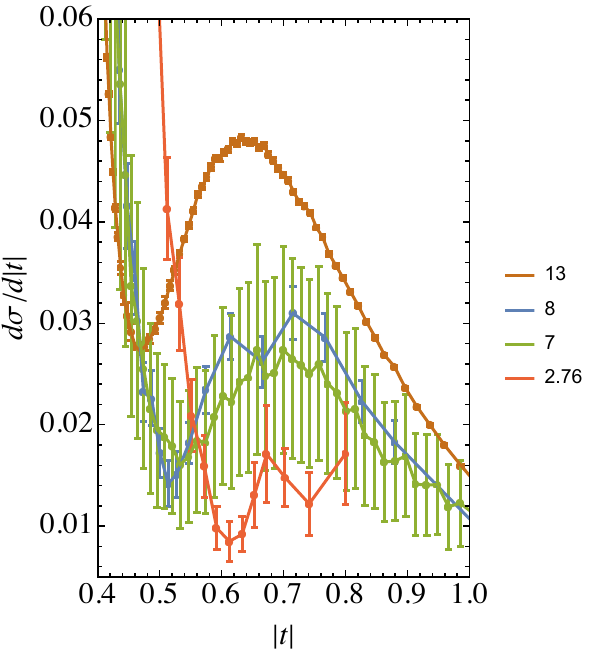}
~\includegraphics[height=6.0cm]{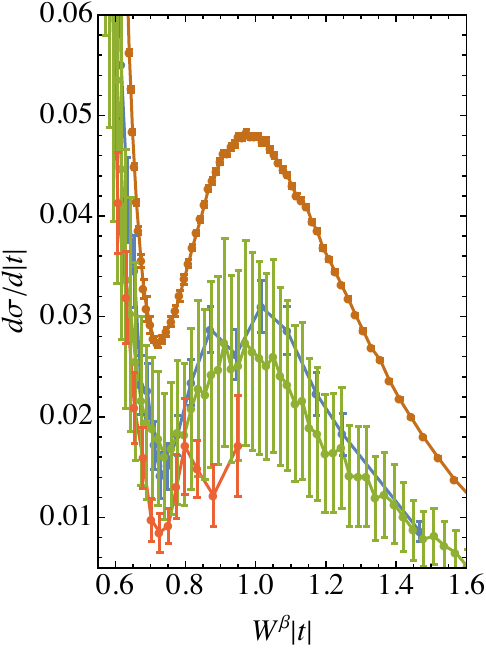}
\caption{Elastic $pp$ cross-section
$d\sigma_{\mathrm{el}}/dt$~[mb/GeV$^{2}$] at the LHC energies  in terms of $|t|$~[GeV$^{2}$] -- left, and in terms of the
scaling variable $W^{\protect\beta}\,|t|$ -- right. We can see that dip and
bumps are aligned after scaling.}%
\label{fig:lhc1}%
\end{figure}

Now, if we plot $d\sigma_{\mathrm{el}}/d|t|$ in
terms of the variable $\tau=(W/(1~\mathrm{TeV}))^{0.1686}\, |t|$, the dip and bump structures
should be aligned independently of the collision energy, as seen in Fig.~\ref{fig:lhc1}, where we show the dip-bump region
for the differential cross-section before and after scaling according to
Eq.~(\ref{eq:dipbumpfits}). Note that power $\beta=0.1686 \pm 0.0027$  is very close to the
energy growth of the total cross-section at the LHC ($0.1739\pm 0.0263$), as can be seen from
Tab.~\ref{tab:sigmas}. 

\begin{figure}[h]
\centering
\includegraphics[height=6.0cm]{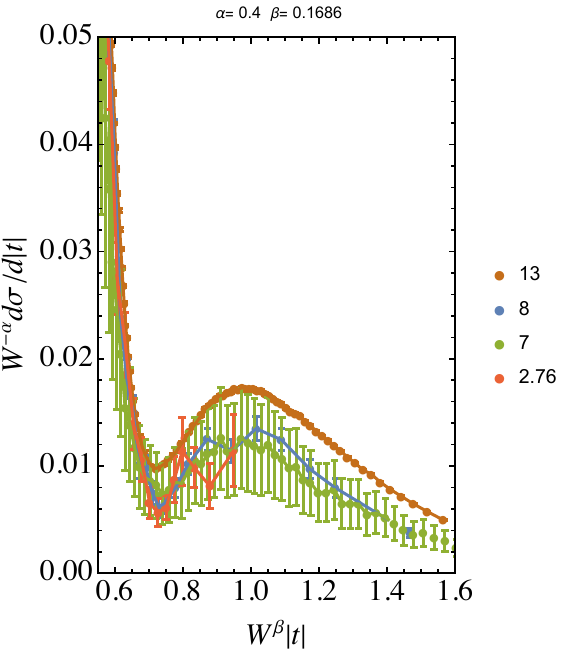}
~\includegraphics[height=6.0cm]{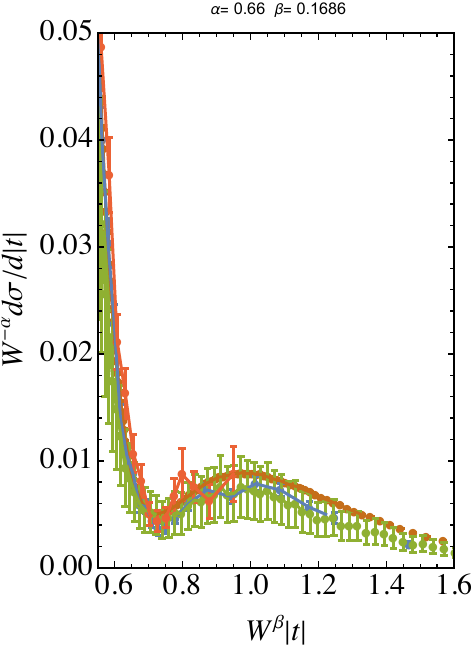}
~\includegraphics[height=6.0cm]{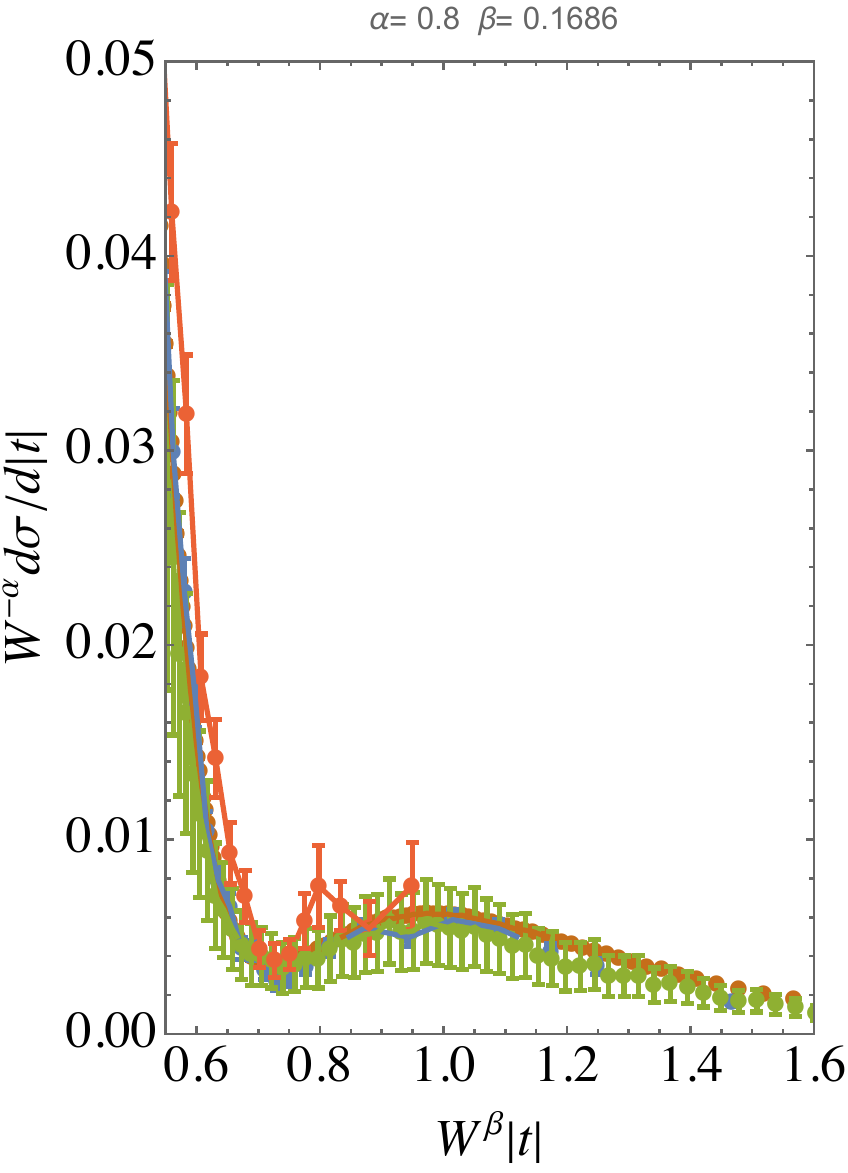}
\caption{Scaled elastic $pp$
cross-section $d\sigma_{\mathrm{el}}/dt$~[mb/GeV$^{2}$] at the LHC energies
 in terms of the scaling variable
$W^{\beta}|t|$ for $\beta=0.1686$ and for $\alpha=0.4$ (left), $\alpha=0.66$ (middle) and
$\alpha=0.8$ (right).}%
\label{fig:lhcfull}%
\end{figure}

Next, we aim at superimposing all curves in the right panel of
Fig.~\ref{fig:lhc1} shifting them vertically by an energy dependent factor. 
To this end we try a simple transformation
(\ref{eq:xsecscaling})
\begin{equation}
\frac{d\sigma_{\mathrm{el}} }{d|t|}(\tau)\rightarrow\left(  \frac
{W}{1\mathrm{TeV}}\right)  ^{-\alpha}\frac{d\sigma_{\mathrm{el}} }{d|t|}%
(\tau)\label{eq:salpha}%
\end{equation}
where the $\alpha$ parameter  has to be extracted from the  data. The result is shown in Fig.~\ref{fig:lhcfull} 
where we have plotted scaled cross-sections (\ref{eq:salpha})
for three values of $\alpha$: 0.4, 0.66 and 0.8.

In the three panels of Fig.~\ref{fig:lhcfull} elastic cross-sections overlap 
or nearly overlap in comparison with the right panel of Fig.~\ref{fig:lhc1}, indicating scaling.
Nevertheless, we definitely need a qualitative measure of GS to find the best value of $\alpha$.
Unfortunately, the LHC data sets are of very different precision, which can be easily seen in the
left panel of Fig.~\ref{fig:lhc1}. Here the 13~TeV data are the best, having the smallest uncertainties, 
the largest density of points and the largest $t$ range. Data points at 7~TeV are also dense 
covering a wide range in of $t$, however they have very large uncertainties. The 8~TeV measurement is
rather dilute (i.e., lower density of points) with moderate  uncertainties, and 2.76~TeV data hardly reach the bump region because of the roman pot detector acceptance.
In this situation\footnote{Note that all ISR data sets have a similar amount of data points in a given range o $|t|$ and commensurate systematic uncertainties.} it is rather
challenging to perform global fits to find the best value of $\alpha$ or both
$\alpha$ and $\beta$ since they would be overconstrained by the 13 TeV data. We investigated such fits over different ranges of $t$, but the results
were not conclusive. Therefore, we have decided to apply the ratio method and
perform $\chi^2$-fit analyses for each data set separately.

\begin{figure}[h]
\centering
\includegraphics[width=6.0cm]{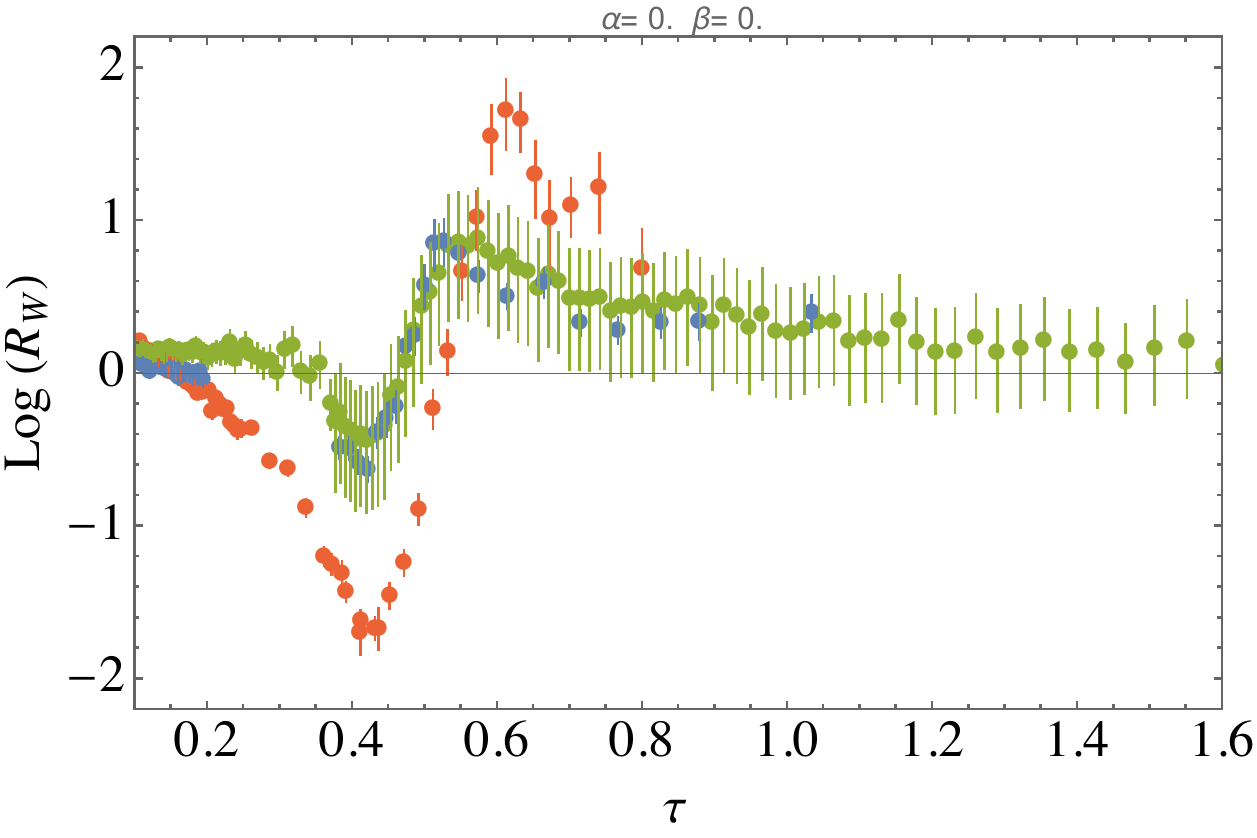}~\includegraphics[width=6.0cm]{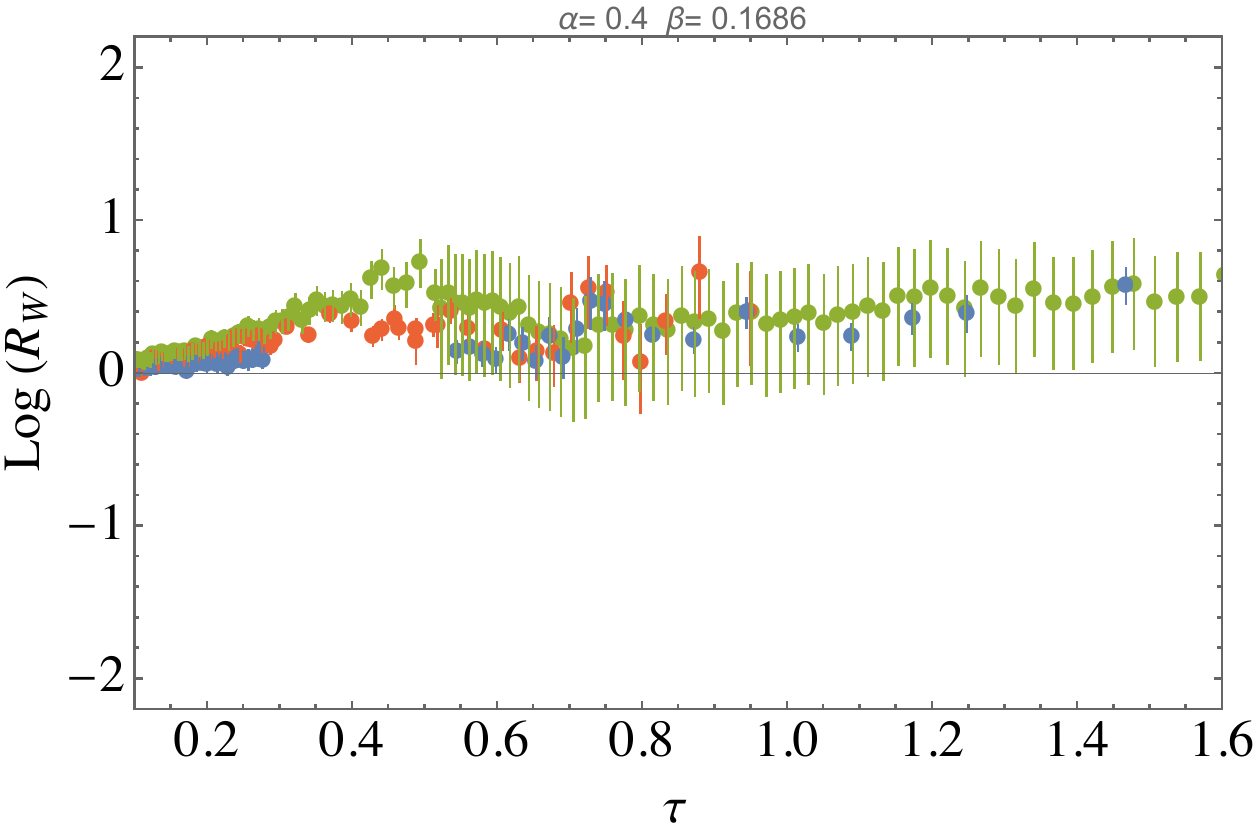}
\vspace{0.2cm}
\includegraphics[width=6.0cm]{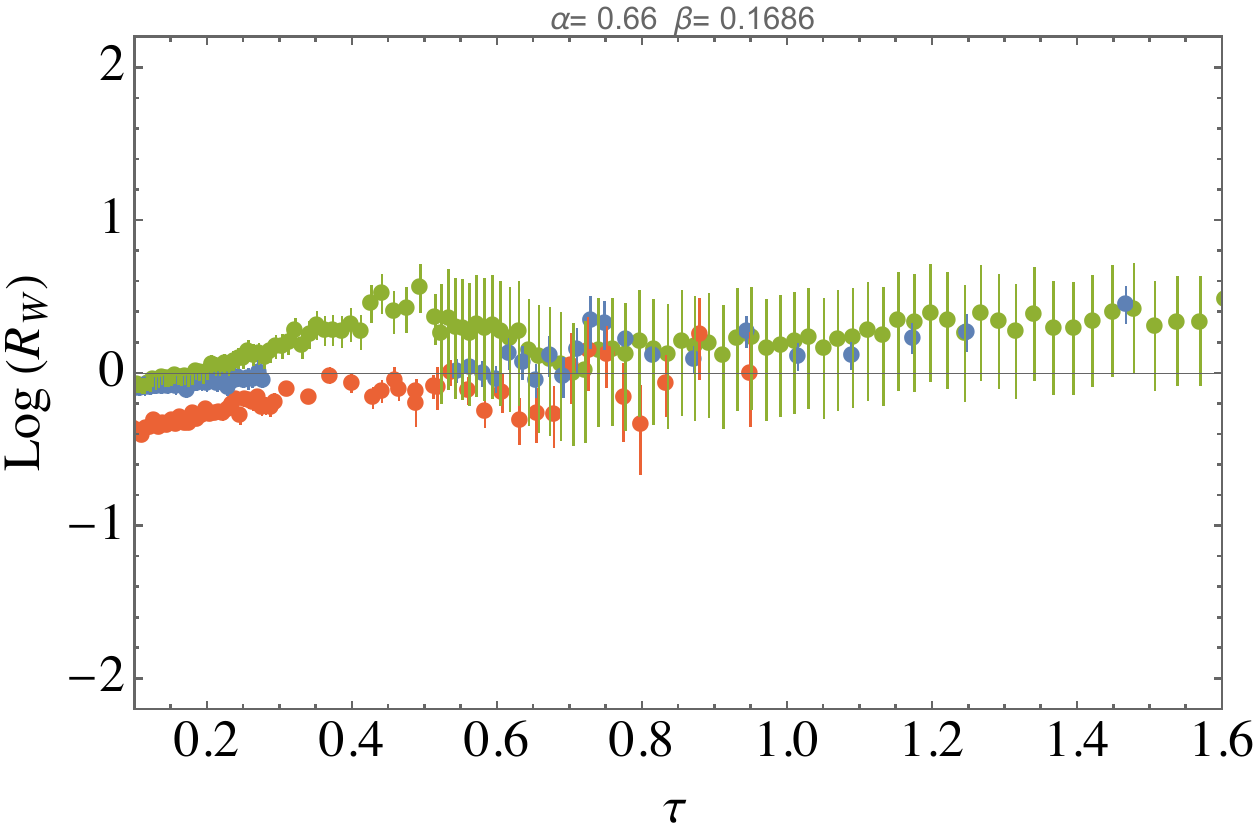}~\includegraphics[width=6.0cm]{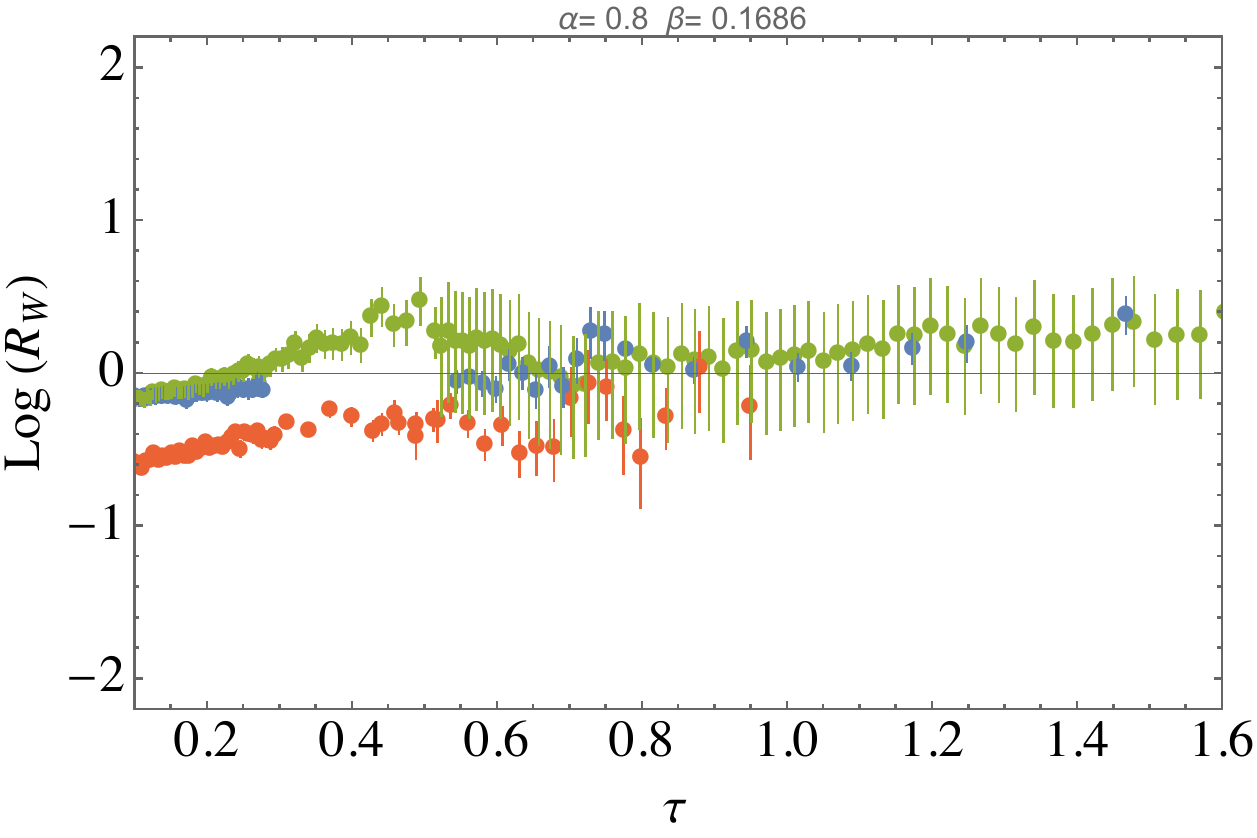}\caption{Ratios
(cf Eqn.\ref{eq:RW}) with respect to $W_{\mathrm{ref}}=13$~TeV for different values of parameter
$\alpha$ and fixed $\beta$ (except for the first panel where no scaling has
been assumed). Blue points correspond to $W=8$~TeV, green to 7~TeV, red to
2.76~TeV. }%
\label{fig:lhcratios}%
\end{figure}

In Fig.~\ref{fig:lhcratios}  we plot ratios (\ref{eq:RW}) for the unscaled
cross-sections and for three different values of $\alpha$ with $\beta$ fixed
to $0.1686$. We have chosen $13$~TeV data as the reference energy. We see that the best
alignment of all ratios is achieved for $\alpha\sim 0.66$, which is somewhat
different than $\alpha=0.61$, the result of Ref.~\cite{Baldenegro:2022xrj}. The reason might be that  Ref.~\cite{Baldenegro:2022xrj} uses a different fitting
strategy, namely the so-called {\it quality factor}. Their goal was to align all points
within certain $t$ range, rather than only the dips and bumps. A small misalignment of the dips and bumps 
can be beneficial for the alignment of the points outside the dip and bump region.

A few comments concerning Fig.~\ref{fig:lhcratios} are in order. Firstly,  we
do not observe any singular behavior of scaled ratios around the dip, which
has been a very pronounced feature at the ISR. Secondly, in contrast to the
ISR, it is impossible to obtain a satisfactory scaling at small $|t|$, and
this is the region where the cross-sections are the largest. Therefore, scaling at the LHC is present only for the values of $|t|$ larger than some cutoff value of the order of 0.4 GeV$^{2}$ outside the Coulomb-nuclear interference region. This means that the integral
(\ref{eq:siginel}) has to be cut at some value of $b$, which may be energy dependent.
This is why we use two different scaling functions $f(s)$ and
$g(s)$, see Eqs.~(\ref{eq:taudef}) and (\ref{eq:xsecscaling}), to get scaling at the LHC, whereas at the ISR these two functions are
related (\ref{eq:BDdDscaling}). 

\begin{figure}[h!]
\centering
\includegraphics[width=9.0cm]{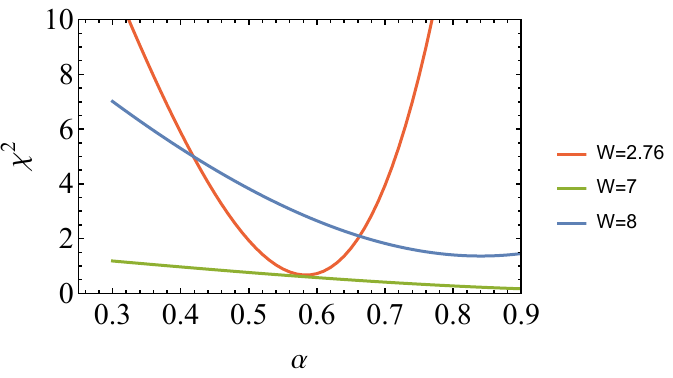}\caption{The values of $\chi^{2}$
(cf Eqn.~\ref{eq:chi2def}) for ratios (cf Eqn.~\ref{eq:RW}) for reference energy
$W_{\mathrm{ref}}=13$~TeV. Blue points correspond to $W=8$~TeV, green to
7~TeV, red to 2.76~TeV.}%
\label{fig:chi2}%
\end{figure}

In order to have a better determination of the scaling exponent $\alpha$ we
plot in Fig.~\ref{fig:chi2} the $\chi^{2}$ values (assuming $\beta= 0.1686$):
\begin{equation}
\chi^{2}(W)=\frac{1}{N_{W}}\sum_{i=1}^{N_{W}}\left(
\frac{R_{W}(\tau_{i})-1}{\delta R_{W}(\tau_{i})}\right)  ^{2}%
\label{eq:chi2def}%
\end{equation}
for the ratios defined in Eq.~(\ref{eq:RW}) in a region $0.35~\mathrm{GeV}%
^{2}< |t| < 1.5~\mathrm{GeV}^{2}$. We see that $\chi^{2}$ scan curve for  $W=8$~TeV
is small and flat, so that the main contribution to the overall
determination of $\alpha$ comes from the limited amount of data at 2.76~TeV, due to the large energy difference between the 2.76 TeV and 13 TeV data sets. We therefore conclude that, 
as already discussed above,
the aggregate analysis of
all data sets together is dominated by the 13 TeV data set. For the purpose of the present
analysis, we have chosen $\alpha=0.66$ where 2.76 and 8~TeV $\chi^2$ cross and are equal approximately to 2.

\section{Other scaling laws}\label{sec:otherScalingLaws}
In Ref.~\cite{Baldenegro:2022xrj} another scaling law for the LHC data has
been proposed
\begin{equation}
\frac{1}{s^{\alpha/2}}\frac{d\sigma_{\mathrm{el}}}{d|t|}(s,t)=\Phi
(\tilde{\tau})\label{eq:BRSscaling}%
\end{equation}
where the scaling variable has been of the form
\begin{equation}
\tilde{\tau}=s^{a} t^{b}\label{eq:BRStau}%
\end{equation}
with the result $\alpha\simeq0.61$ 
and $a \simeq0.065$, $b \simeq0.72$. These
values have been obtained by minimizing the so called {\it quality factor}
(QF) that provides a measure of the alignment of a set of data points along
some universal (yet unknown analytically) curve
\cite{Gelis:2006bs,Beuf:2008mf,Marquet:2006jb}. So, how is this scaling related to
the scaling  discussed here in Sec.~\ref{sec:GS@LHC}? The key
point is that the transformation in Eq.~(\ref{eq:BRStau}) should align the dips (and
bumps) at all LHC energies. This means that the dip (bump) positions
\begin{equation}
\tilde{\tau}_{\mathrm{d}}=s^{a} t_{\mathrm{d}}^{b}=s^{a-b\, \beta/2}B_{\mathrm{dip}%
}^{b}\label{eq:BRStau1}%
\end{equation}
should be energy independent. In the second equality we have used
(\ref{eq:taudef}) with $f(s)=B\, s^{-\beta/2}$. From the energy independence of
(\ref{eq:BRStau1}) we obtain
\begin{equation}
a-b\, \beta/2 =0.\label{eq:zeroabbeta}%
\end{equation}
Substituting $b$ from \cite{Baldenegro:2022xrj} and our $\beta= 0.1686$ we find
that $a=b\, \beta/2 =0.061\pm 0.001$ as compared to $a=0.065$ from the fit of
Ref.~\cite{Baldenegro:2022xrj}, which is only $6$\% off. This small violation
of the constraint (\ref{eq:zeroabbeta}) comes from the minimizing of the QF,
that most probably favors the alignment of the off-dip and off-bump data
points, over the superimposition of dips and bumps themselves.

Furthermore, our result for the power $\alpha=0.66$, which is poorly constrained by the data,
is again in qualitative agreement with \cite{Baldenegro:2022xrj}, where
$\alpha=0.61$.

Obviously Eq.~(\ref{eq:zeroabbeta}) implies the whole family of scaling laws.
This is very much in line with the findings of Ref.~\cite{Baldenegro:2022xrj},
where the scaling has been looked for $\tau$ defined in (\ref{eq:BRStau}) with
$a=A-B$ and $b=1-A$. Equation~(\ref{eq:zeroabbeta}) implies $B=A-\beta
/2(1-A)=A-0.0843(1-A)$, whereas minimization of the QF gives $B=A-0.065$. It is
worth noting, that in the vicinity of the actual best value $A=0.28$, the
present relation reduces to $B=A-\beta/2 \times0.72=A-0.061$.

The best-fit values of $a$ and $b$ satisfying Eq.~(\ref{eq:zeroabbeta}) and the
value of the overall cross-section scaling parameter $\alpha$ are obtained by minimizing the $\chi^{2}$ of the ratios (\ref{eq:RW}) or
by the QF method. However, in view of our analysis of Sec.~\ref{sec:GS@LHC}, 
we would argue that the precision and sample size of the present
data at lower LHC energies is not sufficient for such an analysis to be
meaningful. We only quote the $\chi^{2}$ value for fixed scaling parameters
in Table \ref{tab_chi2}. We see that the robustness of both scaling laws (\ref{eq:taudef}) and (\ref{eq:BRStau}) is compatible.

\begin{table}[h!]
\centering
\begin{tabular}
[c]{|c|c|c|}\hline
$W$ & \multicolumn{2}{c|}{$\chi^{2}$}\\
(TeV) & this paper & Ref.~\cite{Baldenegro:2022xrj}\\
\hline
2.76 & 1.99 & 1.40\\
7 & 0.47 & 0.64\\
8 & 2.11 & 2.96\\\hline
\end{tabular}
\caption{
$\chi^{2}$ for two scalings: present paper with $\beta=0.1686,\,
\alpha=0.66$ and Ref.~\cite{Baldenegro:2022xrj} with
$\alpha=0.61$, $a=0.065$ and $b=0.72$.}%
\label{tab_chi2}%
\end{table}

\section{Dip-bump structure and amplitude parametrization}\label{sec:dipBump}

Let us investigate the consequences of the constant ratio of the bump-to-dip positions on the properties of the scattering amplitudes. We consider 
a typical parametrization of the elastic amplitude used in the literature, which  is a
two component \textit{ansatz} for the rescaled amplitude~\cite{Baldenegro:2022xrj}%
\begin{equation}
\mathcal{A}(s,t)=i\left(  \mathcal{A}_{1}(s,t)+\mathcal{A}_{2}(s,t)e^{i\phi
}\right)
\label{eq:ampls}
\end{equation}
where%
\begin{equation}
\mathcal{A}_{i}(s,t)=N_{i}(s)\,e^{-B_{i}(s)|t|}\label{eq:Ais}%
\end{equation}
with $N_i>0$ and $B_i>0$. The differential cross-section reads then%
\begin{align}
\frac{d\sigma_{\mathrm{el}} }{d|t|}  & =\left\vert \mathcal{A}(s,t)\right\vert
^{2}\nonumber\\
& =\left[  N_{1}^{2}(s)e^{-2B_{1}(s)|t|}+N_{2}^{2}(s)e^{-2B_{2}(s)|t|}%
+2N_{1}(s)N_{2}(s)\cos\phi\,e^{-(B_{1}(s)+B_{2}(s)|t|}\right]  .
\end{align}
We need to find the maximum and minimum of $d\sigma_{\mathrm{el}} /d|t|$
(suppressing for clarity the energy dependence of $N_{1,2}$ and $B_{1,2}$):
\begin{align}
\frac{d}{d|t|}\frac{d\sigma_{\mathrm{el}} }{d|t|}  & =-2N_{1}N_{2}(B_{1}%
+B_{2})e^{-(B_{1}+B_{2})|t|}\nonumber\\
& \left[  \frac{N_{1}B_{1}}{N_{2}(B_{1}+B_{2})}e^{(B_{2}-B_{1})|t|}%
+\frac{N_{2}B_{2}}{N_{1}(B_{1}+B_{2})}e^{-(B_{2}-B_{1})|t|}+\cos\phi\,\right]
\nonumber\\
& \;\nonumber\\
& =0.\label{eq:derzero}%
\end{align}
Therefore at the dip and bump the square bracket $\left[  \ldots\right]  =0$
and we see that $\cos\phi$ must be negative. Let's introduce%
\begin{equation}
\beta(s)=\frac{B_{2}(s)}{B_{1}(s)}>0,\;\nu(s)=\frac{N_{2}(s)}{N_{1}(s)}>0
\end{equation}
and a function%
\begin{equation}
f(s,t)=\frac{N_{1}(s)}{N_{2}(s)}e^{(B_{2}(s)-B_{1}(s))|t|}=\frac{1}{\nu
(s)}e^{(B_{2}(s)-B_{1}(s))|t|}.
\end{equation}
Then Eq.(\ref{eq:derzero}) takes the following form%
\begin{equation}
\frac{1}{1+\beta(s)}f(s,t)+\frac{\beta(s)}{1+\beta(s)}\frac{1}{f(s,t)}%
=-\cos\phi
\end{equation}
and has two solutions for $f$ (suppressing the energy dependence again)%
\begin{equation}
f_{\pm}=\frac{1}{\nu}e^{(B_{2}-B_{1})|t|_{\pm}}=\frac{1}{2}\left[
-(1+\beta)\cos\phi\pm\sqrt{(1+\beta)^{2}\cos^{2}\phi-4\beta}\right]
\end{equation}
and we get%
\begin{equation}
|t|_{\pm}=\frac{1}{B_{2}(s)-B_{1}(s)}\ln\left(  \frac{\nu(s)}{2}\left[
-(1+\beta(s))\cos\phi\pm\sqrt{(1+\beta(s))^{2}\cos^{2}\phi-4\beta(s)}\right]
\right)  .\label{eq:tpm}%
\end{equation}
At this point, we do not know which of the two solutions $|t|_{\pm}$
corresponds to the dip or bump positions; this depends on whether the argument of the logarithm
is smaller or larger than unity. Nevertheless, we see that the bump-to-dip position
ratio is given by%
\begin{equation}
\frac{|t|_{+}}{|t|_{-}}=\frac{\ln\left(  \frac{\nu(s)}{2}\left[
-(1+\beta(s))\cos\phi+\sqrt{(1+\beta(s))^{2}\cos^{2}\phi-4\beta(s)}\right]
\right)  }{\ln\left(  \frac{\nu(s)}{2}\left[  -(1+\beta(s))\cos\phi
-\sqrt{(1+\beta(s))^{2}\cos^{2}\phi-4\beta(s)}\right]  \right)  }
\end{equation}
or by its inverse. For this ratio to be energy independent, the simplest
general solution is
\begin{equation}
\nu(s)=\nu\qquad\mathrm{and}\qquad\beta(s)=\beta,
\end{equation}
i.e., both normalization and slope of two amplitudes (\ref{eq:Ais}) have the
same energy dependence:%
\[
N_{i}(s)=n_{i}N(s)\quad\mathrm{and}\quad B_{i}(s)=b_{i}B(s).
\]

There are three more constraints on the parameters of the amplitudes
(\ref{eq:Ais}). Firstly,
\begin{equation}
(1+\beta)^{2}\cos^{2}\phi-4\beta\geq0,
\end{equation}
which is always satisfied for $\cos\phi=\pm1$ and exlcudes $\cos\phi=0$.
Secondly, because $|t|_{\pm}>0$ both%
\begin{equation}
p_{\pm}\equiv\frac{\nu}{2}\left[  -(1+\beta)\cos\phi\pm\sqrt{(1+\beta)^{2}%
\cos^{2}\phi-4\beta}\right]
\end{equation}
have to be simultaneously smaller or larger than 1. Then%
\begin{equation}
p_{\pm}\gtrless1\rightarrow\beta\gtrless1.
\end{equation}
If $p_{\pm}>1$ then $|t|_{+}=|t_{\mathrm{bump}}|$, and if $p_{\pm}<1$ then
$|t|_{-}=|t_{\mathrm{bump}}|$. It is worth mentioning that the parametrization
of Ref.~\cite{Baldenegro:2022xrj} corresponds to the case $p_{\pm}<1$. Finally, because the positions of the dips are empirically observed to shift towards smaller $|t|$ values with
increasing $s$, $B(s)$ has to be an increasing function of $s$.

\section{Summary and conclusions}\label{sec:summary}

The aim of the present paper was to address two questions: What are the scaling properties
of the elastic $pp$ cross-section at the LHC \cite{Baldenegro:2022xrj}? And 
how is it related to the geometrical scaling
found at the ISR \cite{DiasDeDeus:1973lde,Buras:1973km}?

Our study is based on the key observation that 
the ratio of the bump-to-dip positions ($t_{\mathrm{b}%
}/t_{\mathrm{d}}$) of $d\sigma_{\mathrm{el}}/d|t|$ is constant over three orders of magnitude in energy  (from around 20 GeV at the ISR to 13 TeV at the LHC).  This behavior suggests that there might be a universal behavior of the elastic scattering cross sections as function of a scaling variable $\tau=f(s)|t|$.

It turns out that, despite being at a lower collision energy, the ISR data are well amenable to geometrical scaling because elastic, inelastic, and total cross-sections have the same energy dependence, within the experimental uncertainties. This suggests that the opacity function $\Omega(s,b)=\Omega(b/R(s))$
(\ref{eq:GSdef})
where $R^2(s)\propto \sigma_{\rm inel}(s)$ and the real part of the scattering amplitude encoded in the function $\chi(s,b)$
(\ref{eq:sigmas}) can be neglected. At the ISR  transformation (\ref{eq:tau}) not only superimposes bump
and dip positions at different energies, but also superimposes the elastic cross-sections values over a much wider range of $\tau$, except $\tau_{\rm dip}$
(\ref{eq:BDdDscaling}). This is no longer true at the LHC, which requires rescaling of the cross-section values by another
energy dependent function $g(s)$ what was already observed in Ref.~\cite{Baldenegro:2022xrj}. This means that the scaling
at the LHC differs from the geometrical scaling found at the ISR.

Interestingly, geometrical scaling at the ISR works well even at very small
values of $t$, except for the dip region. The observed violation of geometrical scaling in the ISR data in the dip region was attributed to the vanishment of the imaginary part of the elastic amplitude in
the vicinity of $t=t_{\mathrm{d}}$ \cite{DiasdeDeus:1977af}, 
such violation does not happen at the LHC.

To quantify the scaling properties, we have proposed to use the ratio method
described in Sec.~\ref{sec:GS@ISR}, which allows for the computation of standard $\chi^2$ tests.
We used this method to constrain the scaling function (\ref{eq:xsecscaling})
$g(s)=W^{\alpha}$. The parameter $\alpha$ is primarily constrained by the lowest energy data at 2.76~TeV, which
is not as precise as other measurements at higher energies. Nevertheless, the values of the scaling parameters and
the $\chi^2$ values for the ratios are very close to the results
of Ref.~\cite{Baldenegro:2022xrj} obtained by completely different method 
based on the ``quality factor'' ~\cite{Gelis:2006bs,Beuf:2008mf,Marquet:2006jb}.

Our primary observation that ${\cal T}_{\rm bd}=t_{\mathrm{b}}/t_{\mathrm{d}}={\rm const.}$
and that $t_{\mathrm{b}}=W^{-\beta}$, which for the LHC energies gives
$\beta=0.1686\pm 0.0027$, does not entirely constrain the scaling variable $\tau$.
In fact, any variable $\tilde{\tau}$ of the form $\tilde{\tau}=s^a t^b $  leads to an energy-independent
${\cal T}_{\rm bd}$ if $a-b \beta /2=0$. Therefore, the assumption of energy independence 
of ${\cal T}_{\rm bd}$ leads to a family of scaling laws and the
determination of $a$ (or equivalently of $b$) follows from the alignment
of the points outside of the immediate vicinity of dip and bump regions. In our
case,  $a=\beta/2=0.0843$ and $b=1$, whereas in Ref.~\cite{Baldenegro:2022xrj}
$a=0.065$ and $b=0.72$ leading to a scaling of similar quality. This suggests a flat direction direction in the $a$--$b$ parameter space. More quantitative conclusions are hampered at this point by the limitations on the sample size and precision of the elastic scattering data.

Finally, we have examined the consequences of the energy independence of ${\cal T}_{\rm bd}$ for the parametrizations of the scattering amplitudes, which are customarily chosen as a superposition of two exponential functions (\ref{eq:ampls}). It follows that the energy dependence of both exponents and the normalization factors has to be the same.

\section*{Acknowledgements}

The authors thank R. Peschanski for helpful discussions. CB is supported by the European Research Council consolidator grant no. 101002207.
AMS is  supported by the U.S. Department of Energy grant No. DE-SC-0002145 and within the framework of the Saturated Glue (SURGE) 
Topical Theory Collaboration, as well as  in part by National Science Centre in Poland, grant 2019/33/B/ST2/02588.

\end{document}